\documentclass[sigconf,nonacm]{acmart}

\usepackage{algorithmic}
\usepackage{hyperref}
\usepackage{graphicx}
\usepackage{textcomp}
\usepackage{xcolor}
\usepackage{subcaption}
\usepackage{booktabs}
\usepackage{natbib}
\usepackage{multirow}
\usepackage{balance}
\usepackage{soul, listings,lstautogobble}
\usepackage{ color, colortbl}
\usepackage[normalem]{ulem} %strikethrough
\usepackage{float}
\usepackage{xspace}
\usepackage{geometry}
\usepackage{tikz}
\usepackage{paralist}
\usepackage{dirtree}
\usepackage{mdframed}
\usepackage[most]{tcolorbox}
\usepackage{xcolor}
\ifdefined \NDA
    \usepackage{tex/watermark}
\else
\fi

\usetikzlibrary{arrows.meta, positioning, shapes.geometric, fit, calc, backgrounds}

\definecolor{ListBGColor}{rgb}{0.95,0.95,0.95}
\definecolor{KeywordColor}{rgb}{0,0,0.6}
\definecolor{ListCommentColor}{rgb}{0.4, 0.27, 0.173}
\lstdefinelanguage{console}{
  morekeywords={kerncap, make, python3},
  morecomment=[l]{\#},
  morestring=[b]",
}

\makeatletter
\newcommand\labelline[1]{%
  \def\@currentlabel{\thelstnumber}\label{#1}}
\newcommand\lineref[1]{
  \@ifundefined{r@#1}{0}{\ref{#1}}
}
\makeatother

\AtBeginDocument{%
  \providecommand\BibTeX{{%
    \normalfont B\kern-0.5em{\scshape i\kern-0.25em b}\kern-0.8em\TeX}}}

\begin{document}

%%
%% The "title" command has an optional parameter,
%% allowing the author to define a "short title" to be used in page headers.
%\title{Performance-aware software engineering using LLMs} 
\title{Kerncap: Automated Kernel Extraction and Isolation for AMD GPUs}
%\title{\omniwise: LLM-Based GPU Kernel Performance Prediction}
% \title{Omniwise: LLM-Guided Performance-aware Software Optimization Pipeline}

%%
%% The "author" command and its associated commands are used to define
%% the authors and their affiliations.
%% Of note is the shared affiliation of the first two authors, and the
%% "authornote" and "authornotemark" commands

%% used to denote shared contribution to the research.

\author{Cole Ramos}
\email{Cole.Ramos@amd.com}
\orcid{0009-0005-0007-2893}
\affiliation{%
  \institution{AMD}
  \city{Austin}
  \state{Texas}
  \country{USA}
}

\author{Keith Lowery}
\email{Keith.Lowery@amd.com}
%\orcid{xxxx-xxxx-xxxx-xxxx}
\affiliation{%
  \institution{AMD}
  \city{Austin}
  \state{Texas}
  \country{USA}
}

%%
%% By default, the full list of authors will be used in the page
%% headers. Often, this list is too long, and will overlap
%% other information printed in the page headers. This command allows
%% the author to define a more concise list
%% of authors' names for this purpose.
\renewcommand{\shortauthors}{Ramos, and Lowery}

%%
%% The abstract is a short summary of the work to be presented in the
%% article.
% ============================================================================
% ABSTRACT
% ============================================================================
\begin{abstract}
Iterative GPU kernel tuning is bottlenecked by the scale of the
applications that host the kernels. Rapid iteration requires isolating
the kernel so it can be edited, recompiled, and validated without
rebuilding the full application---but manual isolation requires
reconstructing build flags, dispatch configuration, and runtime inputs
by hand, so developers usually settle for slow in-place edits.

We present \textsc{Kerncap}, an automated kernel extraction tool that
intercepts dispatches at the HSA runtime for both HIP and Triton,
bridging Triton's JIT-only metadata into HSA-level capture via a
lightweight Python compile-hook shim. \textsc{Kerncap} performs an
\emph{address-space closure} of all device memory---a
virtual-address-faithful snapshot that preserves embedded device
pointers without DWARF metadata or pointer chasing---locates kernel
sources, and emits self-contained reproducer projects. HIP reproducers
use a Clang VFS overlay for source-level recompilation without
modifying the original build system; Triton reproducers are
\emph{tuning-pinned}, binding the captured autotuner configuration
into the artifact to preserve the JIT kernel's numerical contract.

\begin{sloppypar}
Across six real-world HIP and Triton workloads spanning traditional
HPC and ML domains on three AMD GPU architectures (CDNA2, CDNA3,
RDNA3), \textsc{Kerncap} extracts and validates kernels from snapshots
ranging from 152~MB to 30~GB---including a VA-faithful capture of
vLLM's Mixture-of-Experts weight pool reached through pointer
indirection. On our llama.cpp case study, \textsc{Kerncap}'s
edit-recompile-validate loop achieves a $13.6\times$ speedup over the
traditional workflow, reducing kernel isolation from a multi-hour
process to a single command. The resulting reproducers also serve as
a substrate for autotuning agents and LLM-driven kernel generators
that need rapid, isolated evaluation of candidates.
\end{sloppypar}
\end{abstract}

%%
%% The code below is generated by the tool at http://dl.acm.org/ccs.cfm.
%% Please copy and paste the code instead of the example below.
%%
%\begin{CCSXML}
%  <ccs2012>
%  <concept>
%  <concept_id>10010147.10010169.10010170.10010174</concept_id>
%  <concept_desc>Computing methodologies~Massively parallel algorithms</concept_desc>
%  <concept_significance>500</concept_significance>
%  </concept>
%  <concept>
%  <concept_id>10003752.10003809.10010031</concept_id>
%  <concept_desc>Theory of computation~Data structures design and analysis</concept_desc>
%  <concept_significance>500</concept_significance>
%  </concept>
%  </ccs2012>
%\end{CCSXML}

%\ccsdesc[500]{Computing methodologies~Massively parallel algorithms}
%\ccsdesc[500]{Theory of computation~Data structures design and analysis}

%%
%% Keywords. The author(s) should pick words that accurately describe
%% the work being presented. Separate the keywords with commas.

%% A "teaser" image appears between the author and affiliation
%% information and the body of the document, and typically spans the
%% page.
% \begin{teaserfigure}
%   \includegraphics[width=\textwidth]{sampleteaser.jpeg}
%   \caption{Seattle Mariners at Spring Training, 2010.}
%   \Description{Enjoying the baseball game from the third-base
%     seats. Ichiro Suzuki preparing to bat.}
%   \label{fig:teaser}
% \end{teaserfigure}

%%
%% This command processes the author and affiliation and title
%% information and builds the first part of the formatted document.
\maketitle
% ============================================================================
% 1. INTRODUCTION
% ============================================================================
\section{Introduction}
\label{sec:introduction}

GPU kernel optimization is most effective when developers can iterate on a 
kernel in isolation: a tight edit-recompile-validate loop is far faster than 
rebuilding and rerunning an entire application. However, in practice, 
isolating a kernel is often the dominant cost in this workflow.
As a concrete data point, on our llama.cpp case study a single full-application
CMake rebuild takes 128~s---roughly $15\times$ the application's 8.3~s baseline
runtime---and this cost is paid on every iteration.

Consider a developer optimizing a single attention kernel within a large LLM 
inference pipeline containing hundreds of kernels. Although the target kernel 
may be small, extracting it requires navigating a multi-thousand-file codebase, 
resolving layers of templates and build system indirection, capturing gigabytes 
of device-resident state, and reconstructing a runnable environment. 
A single mistake---missing a header dependency, misidentifying a dispatch
configuration, or capturing incomplete device memory---invalidates the reproducer.
What should be a fast inner loop instead becomes a brittle, hours-long process.

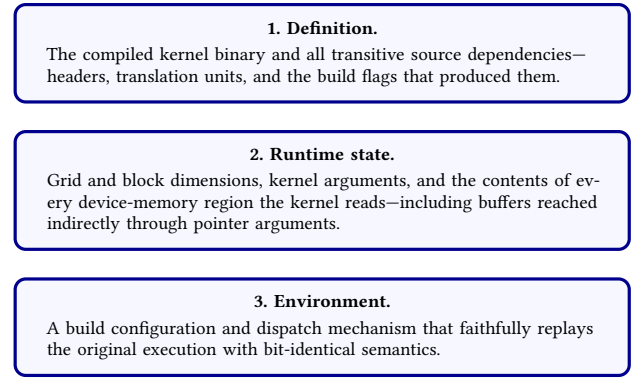
\begin{figure}[H]
\centering
\begin{tikzpicture}[
    font=\footnotesize,
    node distance=0.35cm,
    component/.style={
        draw=blue!55!black, rounded corners=4pt,
        very thick,
        minimum width=\columnwidth-10pt,
        text width=\columnwidth-34pt,
        align=left,
        fill=blue!3,
        inner sep=6pt
    },
    numlabel/.style={font=\bfseries\small, text=blue!70!black},
]
  \node[component] (def) {%
    \makebox[\linewidth][c]{\textbf{1. Definition.}}\\[2pt]
    The compiled kernel binary and all transitive source
    dependencies---headers, translation units, and the build flags
    that produced them.%
  };
  \node[component, below=of def] (state) {%
    \makebox[\linewidth][c]{\textbf{2. Runtime state.}}\\[2pt]
    Grid and block dimensions, kernel arguments, and the contents
    of every device-memory region the kernel reads---including
    buffers reached indirectly through pointer arguments.%
  };
  \node[component, below=of state] (env) {%
    \makebox[\linewidth][c]{\textbf{3. Environment.}}\\[2pt]
    A build configuration and dispatch mechanism that faithfully
    replays the original execution with bit-identical semantics.%
  };
\end{tikzpicture}
\caption{The three components of a GPU kernel reproducer.  Each must
  be recovered in full for a replay to match the original dispatch,
  and the components are tightly coupled---runtime state references
  the definition, and the environment controls how both are
  reassembled.}
\label{fig:three-problems}
\end{figure}

This difficulty arises because a kernel is not a self-contained
artifact.  Reproducing one requires simultaneously recovering three
tightly coupled components, shown in Figure~\ref{fig:three-problems}.
Manual extraction must solve all three at once, which makes the
process labor-intensive and error-prone.

\textsc{Kerncap} automates this process end-to-end by capturing kernels at 
the moment of execution and reconstructing them into standalone, 
self-contained reproducer projects. With a single command:

\begin{lstlisting}[language=console, numbers=none]
kerncap extract attn_fwd \
  --cmd "python bench.py" \
  --source-dir ./src \
  --output ./isolated/attn_fwd
\end{lstlisting}

\noindent
\textsc{Kerncap} profiles the application, intercepts the target kernel dispatch, 
snapshots the complete device memory state, discovers all relevant source files, 
and generates a reproducer with built-in validation to ensure correctness.

\paragraph{Relation to prior work.}
GPU kernel record-and-replay is not a new idea: NVIDIA's Nsight
Compute~\cite{nsight-compute} provides kernel-, range-, and application-level replay modes
for hardware counter collection, and CUPTI's
Checkpoint API (CUDA 11.5+)~\cite{cupti-checkpoint} saves and restores device state across
replay passes.  These tools, however, are
closed-source, target NVIDIA hardware, and treat the captured state as
an internal re-execution buffer rather than as an artifact the
developer can read, edit, and recompile.  \textsc{Kerncap} differs in
three respects: it is open and AMD-native, it produces a self-contained
reproducer that can be edited and rebuilt by the developer, and it
captures the GPU address space \emph{verbatim} rather than via
explicit per-allocation save/restore hooks.  Section~\ref{sec:related}
surveys the full competitive landscape.

\paragraph{Contributions.} Our open-source project\footnote{Kerncap is available at
\url{https://github.com/AMDResearch/intellikit}} currently targets AMD GPUs,
supporting both HIP and Triton kernels. This paper makes the following contributions.

\begin{itemize}
    \item \textbf{Unified HSA-level runtime interception.}  Automated
        kernel dispatch capture at the HSA level for both HIP and
        Triton applications, bridged for Triton by a lightweight
        Python compile-hook shim that supplies the just-in-time
        (JIT)-only metadata (user-visible kernel name, typed
        signature, constexprs) the HSA layer cannot see on its own.
        A single command-line interface (CLI) presents the unified
        capture path to the developer.
    
    \item \textbf{Address-space closure (VA-faithful capture).}  An
        address space is a closure over its pointer graph: by capturing
        every tracked GPU allocation at its original virtual address,
        the embedded pointer relationships (e.g., \texttt{T**}
        arguments) are preserved \emph{for free}, without DWARF
        metadata or pointer chasing.  We realize this with a
        full-address-space snapshot at the HSA level.
    
    \begin{sloppypar}
    \item \textbf{Automated source discovery.}  A multi-phase search
        algorithm that locates kernel definitions, traces
        \texttt{\#include} dependencies, resolves translation units via
        \texttt{compile\_commands.json}, and disambiguates template
        instantiation files using \texttt{nm} symbol lookup.
    \end{sloppypar}

    \item \textbf{Self-contained reproducer generation.}  HIP reproducers
        use a Clang Virtual File System overlay that enables source-level
        recompilation with the exact original compiler command.  Triton
        reproducers are \emph{tuning-pinned}: because JIT kernels carry an
        autotuner-dependent numerical contract that naive capture breaks,
        we bind the captured autotuner configuration into the reproducer
        to preserve it.

    \item \textbf{Validation framework.}  Smoke testing, byte-exact memory
        comparison for HIP variant validation, and tolerance-based
        comparison for Triton reproducers with NaN detection.
    
\end{itemize}
% ============================================================================
% 2. BACKGROUND
% ============================================================================
\section{Background}
\label{sec:background}

\subsection{GPU Kernel Execution on AMD}
\label{sec:bg-execution}

AMD GPUs execute compute kernels through a layered software stack.  At
the lowest level, the \emph{Heterogeneous System Architecture} (HSA)
runtime manages kernel dispatch via AQL (Architected Queuing Language)
packets submitted to hardware queues.  Each kernel dispatch packet
specifies the kernel code object handle, grid and workgroup dimensions,
the kernel argument (\emph{kernarg}) buffer address, and shared memory
requirements.

\begin{sloppypar}
\paragraph{HSA tool interposition.}
AMD's \texttt{ROCProfiler-SDK}~\cite{ROCprofiler-SDK-Documentation}
provides an \texttt{LD\_PRELOAD}-based HSA interception mechanism---a
registration API that exposes the live \texttt{HsaApiTable}, combined
with \texttt{hsa\_amd\_queue\_intercept\_create} for per-packet
callbacks---that tools use to interpose on kernel dispatches and
memory APIs.  \textsc{Kerncap} builds on this mechanism; implementation
details appear in Section~\ref{sec:interception}.

\paragraph{HIP}
HIP (Heterogeneous-Compute Interface for Portability) is AMD's C++
GPU programming model.  Developers write \texttt{\_\_global\_\_}
functions compiled by \texttt{hipcc} (a wrapper around Clang) into
HSACO (HSA Code Object) binaries.  At runtime, HIP loads HSACOs into
HSA executables, resolves kernel symbols, and dispatches them via the
HSA runtime.

\paragraph{Triton.}
OpenAI Triton is a Python-based GPU programming
language \cite{triton}.  Developers write kernel functions decorated with
\texttt{@triton.jit}, which are JIT-compiled to native GPU code.  The
\texttt{@triton.autotune} decorator enables automatic selection among
multiple tile-size configurations by benchmarking each and selecting the
fastest.  Importantly, different tile sizes change the floating-point
accumulation order, causing significant numerical differences in
half-precision computations. This means that a JIT-tuned kernel
carries an implicit \emph{numerical contract} between its tuning
state and its outputs---a contract that any faithful reproducer must
preserve.  

Empirically, on the FP16 \texttt{attn\_fwd} kernel from
\texttt{flash-attn} (B{=}2, H{=}16, S{=}4096, D{=}128) on MI300X,
switching from the fastest autotune config to the second-fastest---only
7.7\% slower---changes 11.3\% of output elements with a maximum
absolute error of $1.22\!\times\!10^{-4}$.  The drift is attributed
to the \texttt{BLOCK\_N} tile, which reorders the
softmax-denominator reduction across the K dimension.
\textsc{Kerncap} preserves the contract by producing
\emph{tuning-pinned} reproducers (Section~\ref{sec:reproducer-gen}).
\end{sloppypar}

\subsection{Manual Kernel Optimization Workflow}
\label{sec:bg-workflow}

Since isolation is prohibitive, developers skip it entirely and 
operate in a coarser loop:

\begin{enumerate}
\begin{sloppypar}
  \item \textbf{Profile}: rank kernels by GPU execution time using
        \texttt{rocprofv3} or similar tools.
  \item \textbf{Search}: locate the kernel's source code by tracing
        mangled names through headers, namespaces, and template
        instantiations---often across thousands of files.
  \item \textbf{Hypothesize}: apply an optimization based on general
        heuristics (e.g., increase tile size, add prefetching, unroll
        an inner loop).
  \item \textbf{Rebuild}: recompile the entire project or supporting
        library, which may take minutes.
  \item \textbf{Evaluate}: re-run the full application and re-profile
        to see if the kernel improved.
  \item \textbf{Repeat}: the optimization often fails to help---or
        makes things worse---because it was not tested against the
        kernel's actual runtime inputs (problem sizes, memory layouts,
        argument values).  The developer reverts and tries again.
\end{sloppypar}
\end{enumerate}

This loop is slow because \emph{every experiment touches the full build
and the full application}.  There is no way to test a kernel change
against the specific dispatch that was profiled without rebuilding and
re-running everything. Writing a standalone harness for the kernel
is possible---and is occasionally done for hot kernels in mature
libraries---but it requires reconstructing the dispatch configuration,
input buffers, and build flags by hand for each new target.  This
up-front cost rarely pays off when the developer's goal is exploratory
optimization across dozens of kernels, so in practice they default to
the outer rebuild-and-rerun loop.
% ============================================================================
% 3. Overview
% ============================================================================
\section{Overview}
\label{sec:overview}

\begin{figure}[H]
\centering
\begin{tikzpicture}[
    font=\footnotesize, % Slightly smaller font for tight columns
    >=Latex,
    node distance=0.7cm, 
    stage/.style={
        draw=blue!55!black,
        rounded corners=3pt,
        very thick,
        minimum width=\columnwidth-10pt, % Fits the column perfectly
        text width=\columnwidth-25pt,
        align=center,
        fill=blue!3,
        inner ysep=5pt
    },
    api/.style={
        font=\sffamily\bfseries,
        text=orange!80!black,
    },
    arrowlabel/.style={
        font=\itshape\small, % Smaller font for connector labels
        text=black!60,
        fill=white,
        inner sep=1pt
    },
    flow/.style={
        draw=black!70,
        thick,
        ->,
    }
]

% --- Nodes ---

% Stage 1
\node[stage] (profile) {
    \textbf{1. Profile}\\[2pt]
    \texttt{rocprofv3 -{}-kernel-trace}
};
\node[api, above=1pt of profile.north west, anchor=south west] {kerncap profile};

% Stage 2
\node[stage, below=1.2cm of profile] (capture) { 
    \textbf{2. Capture runtime state}\\[2pt]
    HIP + Triton: \texttt{LD\_PRELOAD} (rocprofiler-sdk) \\
    Triton: + Python compile-hook shim (metadata)
};

% Stage 3
\node[stage, below=of capture] (source) {
    \textbf{3. Source discovery}\\[2pt]
    HIP: \texttt{compile\_commands.json} \\
    Triton: AST / Import tracing
};

% Stage 4
\node[stage, below=of source] (repro) {
    \textbf{4. Reproducer generation}\\[2pt]
    HIP: Clang VFS \\ Triton: Replay script
};

% Stage 5
\node[stage, below=1.2cm of repro] (validate) { 
    \textbf{5. Replay \& validation}\\[2pt]
    HIP: Byte comp. \\ Triton: \texttt{allclose}
};
\node[api, above=1pt of validate.north west, anchor=south west] {kerncap replay};

% --- Grouping Box (Dotted) ---
\begin{scope}[on background layer]
    \node[
        draw=orange!60!black, dashed, thick, rounded corners=6pt,
        fill=orange!2, inner sep=8pt,
        fit=(capture) (source) (repro),
        label={[api, anchor=south west, xshift=8pt, yshift=4pt]north west:kerncap extract}
    ] (extract-group) {};
\end{scope}

% --- Flow arrows ---
\draw[flow] (profile) -- node[right, arrowlabel] {target kernel} (capture);
\draw[flow] (capture) -- node[right, arrowlabel] {captured state} (source);
\draw[flow] (source) -- node[right, arrowlabel] {source + deps} (repro);
\draw[flow] (repro) -- node[right, arrowlabel] {reproducer} (validate);

\end{tikzpicture}
\vspace{-15pt}
\caption{The \textsc{Kerncap} workflow. The tool abstracts low-level instrumentation into three high-level commands, supporting both HIP and Triton backends.}
\label{fig:pipeline}
\end{figure}
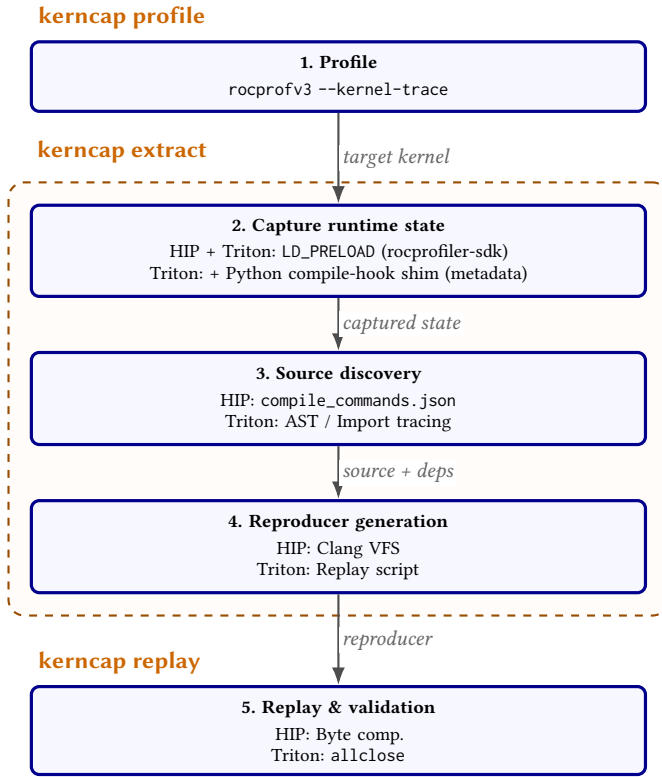

The remainder of this paper presents \textsc{Kerncap}, which automates
the workflow of Section~\ref{sec:bg-workflow} end-to-end.
\textsc{Kerncap} is a command-line tool with a five-stage pipeline.
At the user-facing layer, however, \textsc{Kerncap} exposes only three
commands---\texttt{kerncap profile}, \texttt{kerncap extract}, and
\texttt{kerncap replay}; the five stages describe the internal phases
of an end-to-end run.
Figure~\ref{fig:pipeline} illustrates the pipeline and the artifacts
that flow between stages.

\begin{description}
  \begin{sloppypar}
  \item[\textbf{Profile.}] Ranks kernels by GPU time via
    \texttt{rocprofv3 -{}-kernel-trace -{}-stats}; included for
    workflow completeness, often skipped when the target kernel is
    already known.
  \end{sloppypar}
  \begin{sloppypar}
  \item[\textbf{Capture.}] A shared library (\texttt{libkerncap.so})
    loaded via \texttt{LD\_PRELOAD} intercepts HSA dispatches for
    both HIP and Triton; a lightweight Python compile-hook shim
    supplies the JIT-only metadata (user-visible names, typed
    signatures, constexprs, autotuner configurations) that Triton
    requires, indexed by HSACO SHA-256 in \texttt{name\_map.json}
    (Section~\ref{sec:interception}).
  \end{sloppypar}
  \item[\textbf{Source Discovery.}] Locates the kernel definition and
    its translation unit: HIP via \texttt{compile\_commands.json} and
    \texttt{\#include} tracing, Triton via Python Abstract Syntax
    Tree (AST) walking for
    \texttt{@triton.jit} functions (Section~\ref{sec:source-discovery}).
  \item[\textbf{Reproducer Generation.}] Emits a self-contained
    project with a Clang VFS overlay (HIP) or a tuning-pinned Python
    replay script (Triton), preserving the original build-flag
    fidelity (Section~\ref{sec:reproducer-gen}).
  \begin{sloppypar}
  \item[\textbf{Validation.}] Replays the kernel and compares output:
    byte-exact memory diff for HIP variants, tolerance-based
    \texttt{numpy.allclose} for Triton, or a smoke-test replay when
    no variant is supplied (Section~\ref{sec:replay}).
  \end{sloppypar}
\end{description}

\paragraph{Unified HSA-level capture, language-specific reproducers.}
HIP and Triton kernels operate at fundamentally different abstraction
levels: HIP kernels are compiled ahead-of-time to binary HSACOs and
dispatched via HSA, while Triton kernels exist as Python objects
JIT-compiled at runtime.  \textsc{Kerncap} converges both onto the
same HSA-level capture pipeline---bridging Triton's JIT-only
metadata with a lightweight Python compile-hook shim---and then
diverges back into language-specific reproducer generation (VFS
overlay for HIP, templated Python script for Triton).  The result
is a single \texttt{kerncap extract} CLI command that auto-detects
the kernel language. Detection is a lightweight scan of the user-supplied
\texttt{-{}-source-dir}: if any \texttt{.py} file contains a
\texttt{@triton.jit} or \texttt{@triton.autotune} decorator on a
function whose name matches the target kernel, the Triton path is
selected; otherwise the kernel is treated as HIP.  An explicit
\texttt{-{}-language} user flag can override this heuristic when needed.
% ============================================================================
% 4. DESIGN AND IMPLEMENTATION
% ============================================================================
\section{Design and Implementation}
\label{sec:design}

This section describes the five pipeline stages in detail.  The
implementation is a hybrid C++/Python system: the HSA interception
library and replay binary are written in C++ for direct access to the
HSA API table and low-level memory operations, while the CLI,
source discovery, reproducer generation, Triton capture, and
validation are implemented in Python.

\subsection{Runtime Interception}
\label{sec:interception}

\subsubsection{HIP Path: HSA API Table Hooking}
\begin{sloppypar}
The HIP capture library, \texttt{libkerncap.so}, is loaded via
\texttt{LD\_PRELOAD} and registers with the rocprofiler-sdk framework.
It exports a \texttt{rocprofiler\_configure} entry point that calls
\texttt{rocprofiler\_at\_intercept\_table\_registration} to install a
callback.  When the HSA runtime initializes, this callback receives
the live \texttt{HsaApiTable}---a struct containing function pointers
for the entire HSA API surface.
\end{sloppypar}

\textsc{Kerncap} saves a copy of the original function pointers and
replaces entries in the live table with its own implementations:

\begin{enumerate}
  \item \textbf{Queue interception.}
    \begin{sloppypar}
    \texttt{hsa\_queue\_create} is replaced to call
    \texttt{hsa\_amd\_queue\_intercept\_create}, which creates an intercept queue, followed by
    \texttt{hsa\_amd\_queue\_intercept\_register} to install a
    per-packet callback (\texttt{on\_submit\_packet}).
    \end{sloppypar}

  \item \textbf{Memory tracking.}
    \begin{sloppypar}
    \texttt{hsa\_amd\_memory\_pool\_allocate}, \texttt{hsa\_memory\_allocate}, and the virtual-memory (VMEM) APIs
    (\texttt{hsa\_amd\_vmem\_address\_reserve},
    \texttt{hsa\_amd\_vmem\_map}, and their \texttt{free}/\texttt{unmap}
    counterparts) are hooked to maintain a map from device pointer to
    allocation size.
    \end{sloppypar}

  \item \textbf{Symbol tracking.}
    \begin{sloppypar}
    \texttt{hsa\_executable\_get\_symbol\_by\_name} and
    \texttt{hsa\_executable\_symbol\_get\_info} are hooked to associate
    kernel object handles with their mangled symbol names.
    \end{sloppypar}

  \item \textbf{Code object capture.}
    \begin{sloppypar}
    \texttt{hsa\_code\_object\_reader\_create\_from\_memory} and
    \texttt{hsa\_executable\_load\_agent\_code\_object} are hooked to
    intercept HSACO binary blobs as they are loaded into the runtime.
    The association from kernel object handle to HSACO blob is built
    lazily: when the runtime queries
    \texttt{HSA\_EXECUTABLE\_SYMBOL\_INFO\_KERNEL\_OBJECT}, the tool
    follows the chain \emph{kernel\_object} $\to$ \emph{symbol} $\to$
    \emph{executable} $\to$ \emph{blob}.
    \end{sloppypar}
\end{enumerate}

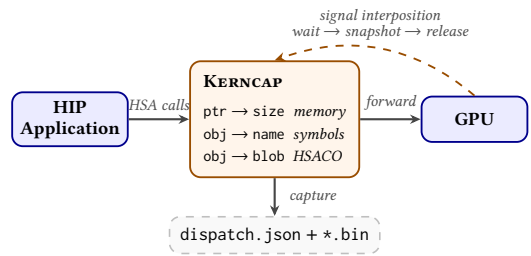
\begin{figure}[H]
\centering
\begin{tikzpicture}[
    node distance=0.4cm and 0.8cm,
    block/.style={
        draw=blue!55!black, rounded corners=4pt,
        minimum height=0.6cm, minimum width=1.4cm,
        align=center, font=\footnotesize\bfseries,
        fill=blue!8, line width=0.7pt
    },
    kcbox/.style={
        draw=orange!60!black, rounded corners=4pt,
        align=left, font=\footnotesize,
        fill=orange!8, line width=0.7pt, inner sep=5pt
    },
    outbox/.style={
        draw=gray!50, rounded corners=4pt, dashed,
        align=center, font=\footnotesize\itshape,
        fill=gray!5, line width=0.6pt, inner sep=4pt
    },
    arrow/.style={
        -{Stealth[length=4pt,width=3.5pt]},
        line width=0.8pt, draw=gray!55!black
    },
    sigarrow/.style={
        -{Stealth[length=4pt,width=3.5pt]},
        line width=0.7pt, draw=orange!60!black, dashed
    },
    annot/.style={font=\scriptsize\itshape, text=gray!60!black},
]

  \node[block] (app) {HIP\\Application};

  \node[kcbox, right=0.8cm of app] (kc) {%
    \textbf{\textsc{Kerncap}}\\[3pt]%
    \scriptsize\texttt{ptr} $\to$ \texttt{size}\enspace\textit{memory}\\[0.5pt]%
    \scriptsize\texttt{obj} $\to$ \texttt{name}\enspace\textit{symbols}\\[0.5pt]%
    \scriptsize\texttt{obj} $\to$ \texttt{blob}\enspace\textit{HSACO}%
  };

  \node[block, right=0.8cm of kc] (gpu) {GPU};

  \node[outbox, below=0.5cm of kc] (out) {%
    \texttt{dispatch.json} $+$ \texttt{*.bin}%
  };

  % Application -> Kerncap
  \draw[arrow] (app) -- (kc)
    node[annot, midway, above] {HSA calls};

  % Kerncap -> GPU (transparent forward)
  \draw[arrow] (kc) -- (gpu)
    node[annot, midway, above] {forward};

  % GPU -> Kerncap (signal interposition).
  % Path travels right-to-left (westward); bend RIGHT = arc northward (above).
  \draw[sigarrow] (gpu.north)
    to[bend right=30]
    node[annot, midway, above, align=center]
      {signal interposition\\[-1pt]wait $\to$ snapshot $\to$ release}
    (kc.north);

  % Kerncap -> captured output
  \draw[arrow] (kc.south) -- (out.north)
    node[annot, midway, right=2pt] {capture};

\end{tikzpicture}
\caption{HSA hook architecture.  \textsc{Kerncap} interposes on the HSA
API table, maintaining three data structures: a pointer-to-size map for
memory tracking, a symbol-to-name map for kernel identification, and a
blob store for HSACO code objects.  All calls are forwarded to the
original HSA implementation.}
\label{fig:hsa-hooks}
\end{figure}

Figure~\ref{fig:hsa-hooks} illustrates the hook architecture.

\paragraph{Signal interposition for post-execution capture.}
When the packet callback identifies a target kernel dispatch, it cannot
simply capture state immediately---the kernel has not yet executed, and
output buffers contain stale data.  \textsc{Kerncap} replaces the
dispatch packet's completion signal with a fresh signal, forwards the
modified packet to the hardware, and blocks on the new signal.  Once
the kernel completes, \textsc{Kerncap} performs the memory snapshot,
then decrements the original signal so the application proceeds
normally.

\subsubsection{Triton Path: HSA-Level Capture with Compile-Hook Shim}
\label{sec:interception-triton}

\begin{sloppypar}
Triton kernels ultimately dispatch through the HSA runtime, so the
same \texttt{libkerncap.so} mechanism that intercepts HIP dispatches
captures Triton dispatches as well.  What HSA does not see is the
Triton-level metadata the reproducer needs: the user-visible kernel
name, the typed Python signature, the values of
\texttt{tl.constexpr} arguments, and the tensor dtypes and strides
that exist only as PyTorch metadata.  We bridge this gap with a
lightweight Python compile-hook shim that runs alongside the HSA
interception path: the shim records Triton-level metadata at
compile time into \texttt{name\_map.json} (indexed by HSACO
SHA-256), and the HSA hook cross-references that file at dispatch
time to reassemble the Triton-level call from the binary kernarg
buffer.
\end{sloppypar}

\paragraph{Compile-hook shim.}
\begin{sloppypar}
A \texttt{sitecustomize.py} module installs a hook on Triton's JIT
compilation path at interpreter startup.  Because Python's
\texttt{site} module imports \texttt{sitecustomize} in every
interpreter, the hook propagates to child processes spawned via
\texttt{multiprocessing.spawn}---including vLLM's EngineCore
workers.  When Triton compiles a kernel, the shim records the
SHA-256 of the emitted HSACO together with the kernel's
user-visible name, typed signature, constexpr bindings, and
per-argument tensor metadata into a shared \texttt{name\_map.json}.
\end{sloppypar}

\paragraph{HSA-level kernarg recovery.}
At dispatch time the HSA hook sees a kernel object handle and a
binary kernarg buffer, but no typed layout.  A C++
kernarg-metadata parser, invoked at executable-load time via
\texttt{amd\_comgr}, reads the AMDGPU code-object metadata blob
and extracts each slot's offset, size, \texttt{value\_kind}, and
type.  Cross-referencing this typed layout against
\texttt{name\_map.json} (indexed by HSACO SHA-256) reassembles the
original Triton call: mangled symbols resolve back to user-visible
names, binary kernarg bytes decode to typed pointer and scalar
arguments, and constexprs recover with their compile-time values.
The resulting \texttt{dispatch.json} and
\texttt{memory\_regions.json} are format-identical to HIP
captures, so the VA-faithful memory snapshot mechanism
(Section~\ref{sec:memory-capture}) applies uniformly.

\paragraph{Autotuner interception.}
When a kernel uses \texttt{@triton.autotune}, the compile-hook
shim also records which keyword arguments originate from
autotuner configurations (e.g., \texttt{BLOCK\_M},
\texttt{num\_warps}) and persists the winning configuration to
\texttt{name\_map.json}, pinning tile sizes, \texttt{num\_warps},
and \texttt{num\_stages} for deterministic replay.

\paragraph{Early host termination.}
For long-running host applications,
completing the full run after the target kernel has already been
captured wastes wall time and risks resource contention with
subsequent \texttt{kerncap} invocations.  Both capture paths
therefore drop a \texttt{capture\_complete} sentinel file as the
final act of artifact serialization.  The \texttt{kerncap extract}
CLI runs a watchdog thread that polls for this sentinel and
\texttt{SIGTERM}/\texttt{SIGKILL}s the child process group the
moment it appears, regardless of whether the host would otherwise
have terminated in seconds or hours.  This is the mechanism behind
the Triton overhead footnote in Table~\ref{tab:overhead}.

\subsection{Memory Capture}
\label{sec:memory-capture}

\subsubsection{HIP Path: VA-Faithful Device Memory Snapshot}

\begin{sloppypar}
After the target kernel completes execution, \textsc{Kerncap} snapshots
\emph{all} tracked device memory regions---not just the kernel's direct
arguments.  The \texttt{snapshot\_all\_tracked\_memory} function
iterates over the pointer-to-size map and streams each region to
\texttt{memory/region\_\{base\_addr\}.bin} via \texttt{hsa\_memory\_copy}
in fixed-size chunks (default 64~MiB, tunable via
\texttt{KERNCAP\_SNAPSHOT\_CHUNK\_BYTES}).  Chunked streaming
bounds the per-region host-memory footprint, which matters for
multi-GiB allocations such as vLLM's KV-cache slabs where a
per-region buffer would otherwise spike host memory consumption.
\end{sloppypar}

A companion file, \texttt{memory\_regions.json}, records each region's
base address, size, allocation type (pool vs.\ VMEM), and whether it
contains the kernarg buffer.  The kernarg segment itself is captured
separately (\texttt{kernarg.bin}) with its exact size queried via
\texttt{HSA\_EXECUTABLE\_SYMBOL\_INFO\_KERNEL\_KERNARG\_SEGMENT\_SIZE}.

\paragraph{Handling embedded device pointers via address-space closure.}
Many GPU kernels use indirect pointers: a kernel argument of type
\texttt{T**} points to a device buffer that itself contains pointers to
other device buffers.  Traditional approaches would require parsing
DWARF debug information or recursively chasing pointers through device
memory.  \textsc{Kerncap} sidesteps this entirely with what we call an
\emph{address-space closure}: an address space is a closure over its
pointer graph, so capturing every tracked allocation at its original
virtual address captures the graph for free.  The replay binary
restores memory at the same virtual addresses, so embedded pointers
remain valid without any interpretation, regardless of how deeply
nested they are.

\paragraph{Bounds of the closure.}
Address-space closure is a device-side guarantee.  Its assumptions
break in three identifiable cases.  (1)~\emph{Host-resident pointers
in kernel arguments}---unified-memory pointers or pointers into
\texttt{mmap}'d host files---are not captured by the device-memory
snapshot and cannot be restored by device VMEM allocation.
(2)~\emph{Address-layout drift across processes}: the closure
guarantees device VAs are reproducible because the replay can request
exact addresses via \texttt{hsa\_amd\_vmem\_address\_reserve}, but
host-side pointers affected by address space layout randomization
(ASLR) and passed in kernargs would not survive a fresh process.  (3)~\emph{Kernel-time pointer arithmetic that
escapes the captured set}---e.g., a kernel that dereferences a
device pointer the runtime never tracked---is by construction outside
the closure.  Section~\ref{sec:limitations} discusses the broader
implications for host-side state.

\paragraph{Module-variable capture.}
\begin{sloppypar}
The pointer-to-size map tracks only allocations routed through
HSA's runtime memory APIs.  It does not cover
\texttt{\_\_constant\_\_} memory populated at executable-load
time via \texttt{hipMemcpyToSymbol}, which appears as
\texttt{HSA\_SYMBOL\_KIND\_VARIABLE} symbols embedded in the
loaded executable.  Portability layers---notably Kokkos's
\texttt{kokkos\_impl\_hip\_constant\_memory\_buffer}---rely on
this path for runtime constants that kernels then dereference,
so ignoring it causes replays to fault on NULL View pointers.
\textsc{Kerncap} enumerates these variable symbols during
capture, reads each one's contents into a per-capture
\texttt{module\_vars/} directory, and restores them in replay
after VMEM allocation but before kernel dispatch.  When a process
loads many executables that export the same symbol name---each
Kokkos kernel carries its own
\texttt{kokkos\_impl\_hip\_constant\_memory\_buffer}---
\textsc{Kerncap} disambiguates the restore by matching the
executable's SHA-256 against the HSACO recorded in the capture,
ensuring the correct blob is restored even when
\texttt{hsa\_executable\_get\_symbol\_by\_name} would otherwise
alias them.
\end{sloppypar}

\paragraph{Ordering and crash safety.}
\textsc{Kerncap} writes \texttt{memory\_regions.json} and
\texttt{dispatch.json} \emph{before} performing the device-to-host
copies.  If an application frees a device buffer between kernel
completion and the snapshot (a race condition inherent to interception
tools), the \texttt{hsa\_memory\_copy} for that region will fail, but
the metadata files and all successfully-copied regions remain intact.
The HSACO binary is also saved before the memory snapshot for the same
reason.

\subsubsection{Triton Path: Unified with HIP}

\begin{sloppypar}
The HSA-based Triton backend
(Section~\ref{sec:interception-triton}) produces
\texttt{memory\_regions.json} and \texttt{region\_\{base\}.bin}
files in the same format as HIP captures, and therefore inherits
the same VA-faithful snapshot, address-space closure, chunked
streaming, and module-variable capture mechanisms described
above. 
\end{sloppypar}

\subsection{Source Discovery}
\label{sec:source-discovery}

\subsubsection{HIP Path: Compile Database and Include Tracking}

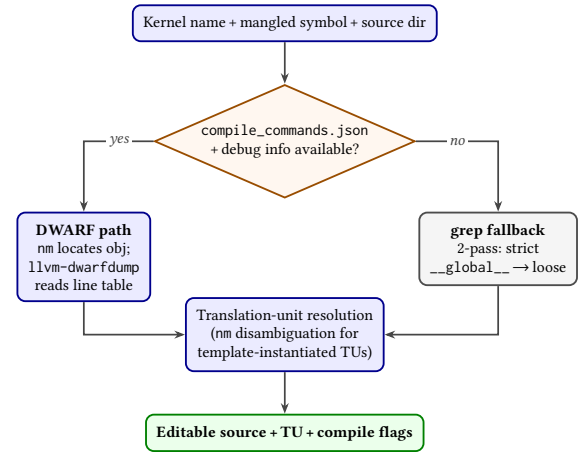
\begin{figure}[H]
\centering
\begin{tikzpicture}[
    node distance=0.55cm and 0.8cm,
    decision/.style={
        draw=orange!60!black, diamond, aspect=2.3,
        align=center, font=\scriptsize,
        fill=orange!6, line width=0.7pt, inner sep=1pt
    },
    primary/.style={
        draw=blue!55!black, rounded corners=3pt,
        align=center, font=\scriptsize,
        fill=blue!8, line width=0.7pt, inner sep=4pt
    },
    fallback/.style={
        draw=gray!60!black, rounded corners=3pt,
        align=center, font=\scriptsize,
        fill=gray!8, line width=0.7pt, inner sep=4pt
    },
    terminal/.style={
        draw=green!50!black, rounded corners=3pt,
        align=center, font=\scriptsize\bfseries,
        fill=green!8, line width=0.7pt, inner sep=4pt
    },
    arrow/.style={-{Stealth[length=4pt,width=3pt]},
                  line width=0.65pt, draw=gray!55!black},
    edgelabel/.style={font=\scriptsize\itshape, text=gray!55!black,
                      fill=white, inner sep=1pt},
]

  \node[primary] (inputs) {Kernel name\,+\,mangled symbol\,+\,source dir};
  \node[decision, below=of inputs] (ccj) {\texttt{compile\_commands.json}\\+\,debug info available?};
  \node[primary, below left=of ccj, xshift=-0.1cm] (dwarf) {\textbf{DWARF path}\\\texttt{nm} locates obj; \\\texttt{llvm-dwarfdump}\\reads line table};
  \node[fallback, below right=of ccj, xshift=0.1cm] (grep) {\textbf{grep fallback}\\2-pass:\ strict\\\texttt{\_\_global\_\_}\,$\to$\,loose};
  \node[primary, below=1.3cm of ccj] (tu) {Translation-unit resolution\\(\texttt{nm} disambiguation for\\template-instantiated TUs)};
  \node[terminal, below=of tu] (done) {Editable source\,+\,TU\,+\,compile flags};

  \draw[arrow] (inputs) -- (ccj);
  \draw[arrow] (ccj.west) -| node[edgelabel, pos=0.25] {yes} (dwarf.north);
  \draw[arrow] (ccj.east) -| node[edgelabel, pos=0.25] {no} (grep.north);
  \draw[arrow] (dwarf.south) |- (tu.west);
  \draw[arrow] (grep.south)  |- (tu.east);
  \draw[arrow] (tu) -- (done);

\end{tikzpicture}
\caption{HIP source-discovery decision tree.  DWARF-based discovery
  is the primary path when debug info and a compile database are both
  available; grep-based search is the fallback.  \texttt{nm}
  disambiguation resolves multi-candidate translation units (e.g.,
  template instantiation files) in both paths.}
\label{fig:source-discovery}
\end{figure}

The HIP source finder takes three inputs---the kernel's demangled
name, its mangled symbol from \texttt{dispatch.json}, and a user
source directory---and resolves the editable source plus its
compile-unit translation unit.  Figure~\ref{fig:source-discovery}
shows the fallback order.

\paragraph{Inputs and base-name extraction.}
\begin{sloppypar}
A demangled C++ kernel name often carries namespaces, template
parameters, and (for truncated profiler output) unbalanced template
brackets.  The \texttt{\_extract\_base\_name} routine strips these to
obtain the unqualified function name used for grep-based matching,
handling cases such as
\texttt{ck::GridwiseGemm<{\ldots}>::Run}\,$\to$\,\texttt{Run} and the
common llama.cpp case \texttt{mul\_mat\_vec\_q<(ggml\_type)39}.
\end{sloppypar}

\paragraph{DWARF-first, grep-fallback.}
\begin{sloppypar}
When a mangled symbol and a \texttt{compile\_commands.json} are both
available, \textsc{Kerncap} runs \texttt{nm} across the listed object
files to locate the one containing the mangled kernel symbol, then
reads that object's DWARF line tables (via \texttt{llvm-dwarfdump},
with a \texttt{readelf} fallback) to extract all source files that
contributed to the translation unit, filtered by the user's
\texttt{-{}-source-dir}.  This path is the only one that works for
framework-generated kernels (e.g., Kokkos) whose
\texttt{\_\_global\_\_} qualifier lives in framework headers rather
than user code~\cite{kokkos}.  If debug info is absent or the
compile database is missing, \textsc{Kerncap} falls back to a
two-pass grep search under the source directory: a strict pass
matching the \texttt{\_\_global\_\_} function \emph{definition} (to
avoid driver files that merely reference the kernel), and a loose
pass on the base name if the strict pass returns nothing.
\texttt{\#include "..."} directives are then recursively traced from
the discovered file up to five levels deep to collect header
dependencies.
\end{sloppypar}

\paragraph{Translation-unit resolution.}
Kernels defined in headers compile through a separate
\texttt{.cu}/\texttt{.cpp} translation unit.  Both discovery paths
resolve this the same way: scan \texttt{compile\_commands.json}
entries whose source \texttt{\#include}s the kernel header (or, if
the database is absent, grep \texttt{.cu} files for the include).
When multiple candidates match---routine for template-instantiation
file sets like llama.cpp's \texttt{mmq-instance-*.cu}---\texttt{nm}
disambiguates by selecting the candidate whose compiled object
contains the exact mangled symbol from \texttt{dispatch.json}.
Compile flags (\texttt{-D}, \texttt{-I}) come from the matched
entry; when no entry is found, \textsc{Kerncap} infers defines by
scanning source files for HIP/ROCm/AMD-guarded \texttt{\#ifdef}
blocks.

\paragraph{Empirical incidence.}
\begin{sloppypar}
On the two HIP workloads in our evaluation (llama.cpp and LAMMPS),
the DWARF path succeeded for both, and \texttt{nm} disambiguation
fired for both (llama.cpp's template-instantiated
\texttt{mul\_mat\_vec\_q} variants and LAMMPS's Kokkos-expanded
\texttt{TagPairEAMKernelC}).  The grep fallback was not load-bearing
in our measurements, but is retained for projects that ship without
debug info or a compile database.
\end{sloppypar}

\subsubsection{Triton Path: AST and Import Tracing}

The Triton source finder operates on Python ASTs:

\begin{enumerate}
\begin{sloppypar}
  \item Parse every \texttt{.py} file in the source directory.
  \item Walk the AST for \texttt{FunctionDef} nodes decorated with
        \texttt{@triton.jit} or \texttt{@triton.autotune}.
  \item Match the kernel name against the function name (supporting
        substring matching for flexibility).
  \item Trace \texttt{ImportFrom} nodes to find helper modules,
        supporting both relative imports (\texttt{from .common import
        ...}) and absolute imports.
  \item Detect package structure (\texttt{\_\_init\_\_.py}) to preserve
        import hierarchies.
\end{sloppypar}
\end{enumerate}

\subsection{Reproducer Generation}
\label{sec:reproducer-gen}

\subsubsection{HIP Reproducers}

\begin{figure}[H]
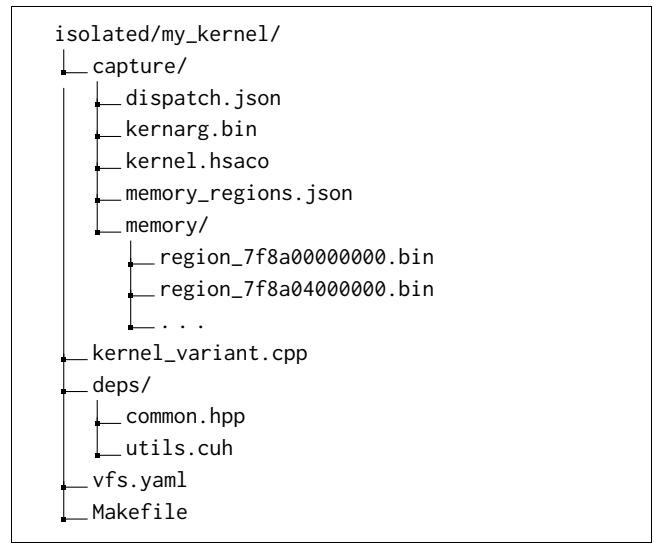

\begin{tcolorbox}[
    width=\columnwidth,       % Force it to fit the column
    sharp corners, 
    colback=white, 
    colframe=black, 
    boxrule=0.5pt,
    left=5pt, right=5pt, top=5pt, bottom=5pt,
    boxsep=0pt                % Reduces extra internal padding
]
\begin{minipage}{\linewidth}
\dirtree{%
.1 isolated/my\_kernel/.
.2 capture/.
.3 dispatch.json.
.3 kernarg.bin.
.3 kernel.hsaco.
.3 memory\_regions.json.
.3 memory/.
.4 region\_7f8a00000000.bin.
.4 region\_7f8a04000000.bin.
.4 \dots.
.2 kernel\_variant.cpp.
.2 deps/.
.3 common.hpp.
.3 utils.cuh.
.2 vfs.yaml.
.2 Makefile.
}
\end{minipage}
\end{tcolorbox}
\caption{Structure of a generated HIP reproducer project.}
\label{fig:reproducer}
\end{figure}

The HIP reproducer generator creates a self-contained project
directory with the structure shown in Figure~\ref{fig:reproducer}.

\noindent
The key components are:

\begin{sloppypar}
\paragraph{Capture data.}
The \texttt{capture/} directory contains the complete VA-faithful
snapshot: dispatch metadata, kernarg buffer, HSACO binary, and all
device memory regions.  The replay binary (\texttt{kerncap-replay})
reads this directory directly.

\paragraph{Editable source.}
The main translation unit is copied as \texttt{kernel\_variant.cpp},
and all traced header dependencies are flattened into \texttt{deps/}
with collision handling (directory-prefixed names when two headers share
a basename).

\paragraph{VFS overlay.}
The Clang Virtual File System overlay (\texttt{vfs.yaml}) is the
mechanism that enables source-level recompilation without modifying
the original build system.  It maps each local file copy back to its
original filesystem path:
\end{sloppypar}

\begin{lstlisting}[numbers=none, basicstyle=\ttfamily\scriptsize]
{"version": 0, "roots": [
  {"type": "directory",
   "name": "/original/src/kernels",
   "contents": [
     {"type": "file",
      "name": "gemm.cuh",
      "external-contents": "/isolated/deps/gemm.cuh"}
   ]}
]}
\end{lstlisting}

\noindent
When the Makefile's \texttt{recompile} target runs, it invokes the
\emph{exact original compiler command} from
\texttt{compile\_commands.json} with three additional flags:
\texttt{-ivfsoverlay vfs.yaml} (to substitute edited files),
\texttt{-{}-cuda-device-only} and \texttt{-{}-no-gpu-bundle-output} (to
produce a raw HSACO instead of a fat binary).  This ensures 100\%
flag and dependency fidelity: the recompiled HSACO uses the same
optimization level, architecture target, and include paths as the
original build.

\paragraph{Makefile.}
The generated Makefile provides four key targets:
\texttt{run} (replay the captured kernel),
\texttt{recompile} (rebuild the HSACO from edited source),
\texttt{run-variant} (replay with the recompiled HSACO), and
\texttt{validate-variant} (compare variant output against baseline).

\subsubsection{Triton Reproducers}

\begin{sloppypar}
Triton reproducers are generated from a Jinja2 template that produces a
standalone \texttt{reproducer.py} script.  The script:

\begin{enumerate}
  \item Loads captured tensor data from binary files using
        \texttt{numpy.fromfile}, converting to the original torch dtype
        and shape.
  \item Imports the kernel function from the copied source module.
  \item Launches the kernel with the captured grid dimensions and
        arguments.
  \item Saves output tensors for validation.
\end{enumerate}

\paragraph{Tuning-pinned reproducers.}
A JIT-compiled Triton kernel carries an implicit numerical contract
between its tuning state (tile sizes, warp/stage counts) and its
outputs (Section~\ref{sec:bg-execution}).  Naive replay breaks this
contract: re-running an extracted kernel under the autotuner can
silently select a different configuration and produce numerically
different outputs.  We preserve the contract by producing
\emph{tuning-pinned reproducers}.  If the captured kernel used
\texttt{@triton.autotune}, the reproducer bypasses the autotuner
entirely; instead of calling the autotuner wrapper, it invokes
\texttt{kernel.fn[grid](**config)} directly with the captured winning
configuration (tile sizes, \texttt{num\_warps}, \texttt{num\_stages}),
binding the tuning state into the reproducer artifact.

\paragraph{Package-aware source copying.}
If the kernel module lives inside a Python package (directory contains
\texttt{\_\_init\_\_.py}), the generator copies the entire package
directory rather than individual files, preserving relative import
chains.
\end{sloppypar}

\subsection{Replay and Validation}
\label{sec:replay}

\subsubsection{VA-Faithful HSA Replay}

The replay binary (\texttt{kerncap-replay}) restores the captured
device memory state and re-dispatches the kernel using raw HSA APIs,
entirely bypassing the HIP runtime.  The replay proceeds in six stages:

\begin{enumerate}
\begin{sloppypar}
  \item \textbf{Parse metadata.}  Read
        \texttt{memory\_regions.json} to determine the set of virtual
        address ranges that must be restored, and
        \texttt{dispatch.json} for grid dimensions, kernarg size, and
        the mangled kernel symbol name.

  \item \textbf{Pre-mmap.}  Before \texttt{hsa\_init()}, issue
        \texttt{mmap(MAP\_FIXED\_NOREPLACE)} calls for each captured
        region.  This reserves the virtual address ranges so the HSA
        runtime's SVM (Shared Virtual Memory) aperture initialization
        does not claim them.

  \item \textbf{HSA initialization.}  Call \texttt{hsa\_init()}, then
        \texttt{munmap} the pre-reserved ranges.

  \item \textbf{VMEM allocation.}  For each region, call
        \texttt{hsa\_amd\_vmem\_address\_reserve} with the exact
        captured base address, create a backing handle via
        \texttt{hsa\_amd\_vmem\_handle\_create}, and map it with
        \texttt{hsa\_amd\_vmem\_map}.  If the reserve returns a
        different address than requested, the replay aborts---VA
        faithfulness is a hard requirement.

  \item \textbf{Memory restore and dispatch.}  Copy captured data into
        each restored region.  Load the HSACO (or an override via
        \texttt{-{}-hsaco}), resolve the kernel symbol by mangled name,
        allocate a kernarg buffer, fill it from \texttt{kernarg.bin},
        and submit an AQL dispatch packet.

  \item \textbf{Output dump.}  If \texttt{-{}-dump-output} is specified,
        copy all regions back to host after kernel completion and write
        them to \texttt{output/region\_\{base\}.bin} for comparison.
\end{sloppypar}
\end{enumerate}

The replay supports multi-iteration benchmarking with optional
memory re-copy between iterations (\texttt{-{}-no-recopy} for stateful
measurement).

\paragraph{Robustness.}
We have validated the replay pipeline on ROCm~$\geq$~7.0, where the
rocprofiler-sdk interception framework is stable.  The Stage~3
relocation-abort path was not observed to trigger across our
evaluation.%
{}
Should a future driver invalidate the pre-reservation assumption,
the replay aborts loudly with a descriptive error rather than
silently corrupting the captured pointer graph.

\subsubsection{Replay Validation}

The validator implements three comparison strategies:

\paragraph{Smoke test (HIP baseline).}
When no variant HSACO is provided, validation simply confirms that
the captured kernel replays without crashing.  This serves as a basic
sanity check that the capture data is complete and the address space
was restored correctly.

\paragraph{Byte-exact comparison (HIP variant).}
When \texttt{-{}-hsaco} provides a recompiled HSACO, the validator runs
two full replays: one with the captured HSACO (baseline) and one with
the variant HSACO.  Both replays use \texttt{-{}-dump-output} to snapshot
post-execution memory.  The validator then compares every output region
byte-for-byte, reporting the number of differing bytes and their
percentage of total region size.

\paragraph{Tolerance-based comparison (Triton).}
For Triton reproducers, the validator runs \texttt{reproducer.py} and
compares the output tensors against captured reference data using
\texttt{numpy.allclose} with configurable absolute (\texttt{atol}) and
relative (\texttt{rtol}) tolerances.  The validator detects and reports
NaN values explicitly, noting that they typically indicate uninitialized
device memory, half-precision overflow, or buffer size
misinterpretation.

\paragraph{Closing the loop.}
Returning to Figure~\ref{fig:three-problems}: the \emph{definition}
is recovered by HSACO capture (Section~\ref{sec:interception})
together with DWARF-based source discovery
(Section~\ref{sec:source-discovery}) and VFS-flattened reproducer
assembly (Section~\ref{sec:reproducer-gen}); the
\emph{runtime state} by address-space closure---the VA-faithful
device-memory snapshot that preserves embedded pointer graphs
without DWARF or pointer chasing
(Section~\ref{sec:memory-capture}); and the \emph{environment} by
exact-flag single-file recompile via Clang VFS
(Section~\ref{sec:reproducer-gen}) and VA-faithful HSA replay
(Section~\ref{sec:replay}).  The tight coupling between
components---particularly between runtime-state capture and
environment replay---is why \textsc{Kerncap} treats capture and
replay as a single mechanism rather than two independent tools.
% ============================================================================
% 5. RESULTS
% ============================================================================
\section{Results}
\label{sec:results}

We evaluate \textsc{Kerncap} along three research questions:

\begin{itemize}
  \item \textbf{RQ1 (Correctness and breadth).}  Does \textsc{Kerncap}
    successfully extract and validate kernels across a diverse set of
    real-world workloads, spanning both HIP and Triton and varying
    application complexity?
  \item \textbf{RQ2 (Overhead).}  What is the wall-clock cost imposed
    on the host application during capture, and how does it decompose
    into per-dispatch interception cost versus one-time memory snapshot
    cost?
  \item \textbf{RQ3 (Iteration speedup).}  How much does the isolated
    edit-recompile-validate loop accelerate kernel optimization
    compared to the traditional full-application workflow?
\end{itemize}

% ----------------------------------------------------------------------------
\subsection{Experimental Setup}
\label{sec:setup}

\begin{sloppypar}
We evaluate \textsc{Kerncap} across three AMD GPU architectures spanning
AMD's data-center (CDNA) and consumer (RDNA) GPU families: AMD
Instinct\texttrademark{} MI300X (gfx942, CDNA3),
AMD Instinct\texttrademark{} MI210 (gfx90a, CDNA2), and AMD Radeon\texttrademark{}
PRO~W7900 (gfx1100, RDNA3).  The CDNA hosts use AMD EPYC\texttrademark{}
9684X CPUs; the RDNA host uses an AMD Ryzen\texttrademark{} Threadripper
PRO~5975WX.  All systems run ROCm~7.2.0 on RHEL~9.6.
Table~\ref{tab:workloads} summarizes the six benchmark workloads.
\end{sloppypar}

\begin{table}[H]
\centering
\caption{Benchmark workloads used for evaluation, spanning three
  orthogonal axes: compiled vs.\ JIT (HIP, Triton, Inductor),
  authored vs.\ framework-generated (hand-written, Kokkos, Tensile,
  Inductor), and snapshot sizes from 152~MB to 30~GB.}
\label{tab:workloads}
\footnotesize
\setlength{\tabcolsep}{4pt}
\begin{tabular}{@{}ll >{\ttfamily}l >{\raggedright\arraybackslash}p{2.9cm}@{}}
\toprule
\textbf{Workload} & \textbf{Lang.} & \textrm{\textbf{Target kernel}} & \textbf{Driver / Input} \\
\midrule
llama.cpp
  & HIP
  & mul\_mat\_vec\_q
  & \texttt{llama-bench}, GPT-OSS 20B MXFP4 \\[3pt]
LAMMPS
  & HIP
  & TagPairEAMKernelC
  & \texttt{lmp} (Kokkos), EAM, $8^3$ grid \\[3pt]
rocBLAS GEMM
  & HIP
  & Cijk\_Ailk\_Bljk\_HHS\_BH
  & \texttt{rocblas-bench}, sgemm $4096^3$ \\[3pt]
Flash Att.\,2
  & Triton
  & attn\_fwd
  & \texttt{sample\_attn\_layer.py}, FP16 \\[3pt]
vLLM
  & Triton
  & fused\_moe\_kernel
  & \texttt{vllm bench latency}, Qwen1.5-MoE-A2.7B \\[3pt]
torch.compile
  & Triton
  & \rmfamily triton\_poi\_fused\_relu\_0
  & \texttt{torch.compile} micro-bench, fp16 matmul \\
\bottomrule
\end{tabular}
\end{table}

\noindent
The six workloads span three orthogonal axes that
the design space requires us to defend: \emph{compiled vs.\ JIT} (HIP
ahead-of-time vs.\ Triton runtime vs.\ torch.compile's Inductor backend
on-demand); \emph{authored vs.\ framework-generated} (hand-written
CUDA-style kernels vs.\ Kokkos template instantiation vs.\
Tensile/Inductor codegen); and \emph{small vs.\ large memory footprint}
(152~MB to 30~GB snapshots).  Two workloads
(rocBLAS GEMM, torch.compile) deliberately have no recoverable source,
exercising the HSACO-only fallback path described in
Section~\ref{sec:source-discovery}.

% ----------------------------------------------------------------------------
\subsection{Extraction and Validation}
\label{sec:results-extraction}

For each workload we run the full \textsc{Kerncap} pipeline
(\texttt{kerncap extract} followed by \texttt{kerncap validate}) on
each available architecture and record the number of device memory
regions captured, the total snapshot size, and the end-to-end
extraction wall-clock time.  Table~\ref{tab:extraction} presents the
results across the three GPUs.  Source-file discovery is workload-stable
(6~files for llama.cpp, 10--11 for LAMMPS, 2 for the hand-authored
Triton workloads, 0 for the no-source workloads) and is omitted from
the table for compactness.

\begin{table}[H]
\centering
\caption{Cross-architecture extraction results.}
\label{tab:extraction}
\footnotesize
\setlength{\tabcolsep}{4pt}
\begin{tabular}{@{}llccc@{}}
\toprule
\textbf{Workload} & \textbf{Arch}
  & \textbf{Reg.}
  & \textbf{Snap.\,(MB)}
  & \textbf{Time (s)} \\
\midrule
\multirow{3}{*}{llama.cpp}
  & gfx942  & 21  & 12{,}233 & 15.6 \\
  & gfx90a  & 21  & 12{,}126 & 20.9 \\
  & gfx1100 & 20  & 11{,}520 & 18.8 \\
\midrule
\multirow{3}{*}{LAMMPS}
  & gfx942  & 84  & 8{,}449  & 25.0 \\
  & gfx90a  & 84  & 8{,}421  & 28.2 \\
  & gfx1100 & 83  & 8{,}415  & 47.7 \\
\midrule
\multirow{3}{*}{rocBLAS GEMM}
  & gfx942  & 10  & 248      & 9.0  \\
  & gfx90a  & 10  & 152      & 11.2 \\
  & gfx1100 & 10  & 152      & 4.4  \\
\midrule
\multirow{2}{*}{Flash Att.\,2}
  & gfx942  & 13  & 178      & 9.1  \\
  & gfx90a  & 13  & 178      & 13.2 \\
\midrule
\multirow{2}{*}{vLLM}
  & gfx942  & 185 & 30{,}074 & 60.9 \\
  & gfx90a  & 186 & 29{,}301 & 85.2 \\
\midrule
\multirow{2}{*}{torch.compile}
  & gfx942  & 12  & 227      & 9.4  \\
  & gfx90a  & 12  & 227      & 13.5 \\
\bottomrule
\end{tabular}

\smallskip\scriptsize
\emph{Reg.}~=~device memory regions captured;
\emph{Snap.}~=~total snapshot size;
\emph{Time}~=~end-to-end wall-clock cost of \texttt{kerncap extract},
including post-capture reproducer generation.  The instrumented
host-app run alone is reported in Table~\ref{tab:overhead};
the difference (largest for LAMMPS at ${\sim}10$~s) is dominated by
source-discovery and reproducer-assembly cost.
Triton workloads tested on gfx942 and gfx90a only.
\end{table}

\textsc{Kerncap} extracts and validates kernels from every workload on
every architecture for which it was tested.  Run-to-run drift in the
captured-region count (e.g., LAMMPS at 84~vs.~83 regions across archs)
reflects natural variation in the Kokkos memory pool's slab
allocations between runs, not a capture-path discrepancy.  Snapshot
sizes are otherwise stable across architectures within a few percent,
confirming that the HSA-level capture path enumerates the same VAs
regardless of underlying ISA family.

\paragraph{HIP validation.}
The llama.cpp reproducer passes full validation on all three
architectures: the captured kernel replays without error, and
recompiling the reproducer's source without modification
(\texttt{make recompile}) produces byte-identical output across all
21~compared memory regions on every arch.  This confirms end-to-end
source-level fidelity through the VFS overlay mechanism on both
CDNA and RDNA targets~\cite{llamacpp}.  On gfx1100, the kernel
replays correctly with the wave32 codegen path
(\texttt{Block:~32$\times$1$\times$1}, vs.\ wave64's
\texttt{32$\times$2} on CDNA), demonstrating that the capture
metadata and replay harness are wavefront-width agnostic.
LAMMPS similarly passes full validation across all three
architectures: the isolated \texttt{TagPairEAMKernelC} kernel
replays byte-identically with 83--84~restored memory regions
spanning ${\sim}$8.4~GB of device state per arch~\cite{LAMMPS}.

\paragraph{HSACO-only path (no source).}
Two workloads---rocBLAS GEMM (Tensile codegen) and torch.compile
(Inductor JIT)---deliberately exercise the HSACO-only reproducer
path, in which source recovery is out of scope because
the kernel's source representation is either an internal codegen
template (Tensile) or a runtime-emitted JIT artifact that does not persist in a
stable on-disk location (Inductor).  In both cases, \textsc{Kerncap} gracefully degrades:
the captured HSACO replays correctly under \texttt{kerncap replay},
and \texttt{kerncap validate} confirms bit-exact replay against the
captured baseline.  The \texttt{make recompile} edit-loop is
unavailable for these workloads by design, but the reproducer remains
useful for hardware-counter profiling, autotuner experimentation, and
input-perturbation studies---capabilities no other tool currently
provides for Tensile or Inductor-emitted kernels.  On gfx1100,
rocBLAS GEMM extracts and validates in 4.4~s, confirming that the
HSACO-only path is portable across ISA families.

\paragraph{Triton validation.}
\begin{sloppypar}
Both hand-authored Triton workloads (Flash~Attention and vLLM) pass
element-wise validation with zero error
(\texttt{max\_error=0.0}) on the architectures tested.  The
reproducers pin the autotuner configuration captured at runtime,
bypassing re-tuning and ensuring deterministic
replay~\cite{flashattention2,vllm}.  vLLM's
\texttt{fused\_moe\_kernel} is a noteworthy demonstration
of VA-faithful capture: the kernel dereferences indirection layers
(the per-expert weight tables in the MoE routing path), and
\textsc{Kerncap}'s HSA-based backend captures the full device-memory
closure those pointers reach---185 regions totaling
${\sim}$30~GB, dominated by 24$\times$702~MB gate/up-projection
tensors and 24$\times$330~MB down-projection tensors
(${\sim}$24~GB of expert weights, ${\sim}$5~GB
activations/workspace)---enabling bit-exact replay of kernels that
argument-only capture cannot reproduce.
\end{sloppypar}

% ----------------------------------------------------------------------------
\subsection{Capture Overhead}
\label{sec:results-overhead}

A na\"ive overhead measurement---comparing total wall-clock time with
and without instrumentation---conflates two fundamentally different
costs: (1)~the \emph{interception tax}, a continuous per-dispatch
callback cost that scales with kernel count, and (2)~the
\emph{capture cost}, a one-time device-memory snapshot that scales
with memory footprint, not application duration.  Reporting a single
overhead ratio lets workload duration confound the result: for
comparable capture costs ($\sim$7--14~s on llama.cpp across archs), a
10-second baseline yields a $1.9$--$2.6\times$ wall-clock ratio, while
a multi-minute workload such as vLLM sees the same fixed cost
amortized to roughly $11\%$ overhead.

We therefore decompose \textsc{Kerncap}'s overhead into these two
components.  For each configuration we perform $N{=}10$ measured runs
after 1~warmup run (discarded) and report median~$\pm$~std.\ dev.
For HIP workloads, we measure an \emph{interception-only}
configuration by loading \texttt{libkerncap.so} via
\texttt{LD\_PRELOAD} while targeting a nonexistent kernel: all HSA
hooks and per-dispatch callbacks execute, but no capture occurs.
The \emph{full capture} configuration targets the real kernel,
triggering the one-time memory snapshot.
For Triton workloads, the Python-level \texttt{JITFunction.run}
monkey-patch adds negligible per-call overhead (a string comparison)
without capture, so interception-only is omitted.
Table~\ref{tab:overhead} presents the wall-clock decomposition.

\begin{table}[H]
\centering
\caption{Wall-clock overhead decomposition on gfx942 ($N{=}10$,
  1~warmup). \emph{Intercept.}\ is the
  ratio of interception-only wall time to \emph{Base}.
  HIP \emph{Full cap.}\ runs the host application to completion
  whereas Triton terminates once the target dispatch is
  captured.  \emph{Full cap.}\ measures the instrumented host-app
  run only; end-to-end \texttt{kerncap extract} time including
  reproducer assembly is reported in Table~\ref{tab:extraction}.}
\label{tab:overhead}
\footnotesize
\setlength{\tabcolsep}{3pt}
\begin{tabular}{@{}lccccc@{}}
\toprule
\textbf{Workload}
  & \textbf{Base}
  & \textbf{Intercept.}
  & \textbf{Full cap.}
  & \textbf{Snap.}
  & \textbf{Cap.} \\[-2pt]
  & \textbf{(s)}
  & \textbf{ratio}
  & \textbf{(s)}
  & \textbf{(MB)}
  & \textbf{cost (s)} \\
\midrule
\multicolumn{6}{@{}l}{\textbf{\textit{HIP}} \emph{(LD\_PRELOAD instrumentation)}} \\[2pt]
llama.cpp     & $8.30 \pm 0.04$  & $1.06\times$ & 15.8 & 12{,}233 & 7.0 \\
LAMMPS        & $9.96 \pm 0.02$  & $1.04\times$ & 15.3 & 8{,}449  & 4.9 \\
rocBLAS GEMM  & $3.90 \pm 0.03$  & $2.04\times$ & 8.2  & 248      & 0.2 \\
\midrule
\multicolumn{6}{@{}l}{\textbf{\textit{Triton}} \emph{(Python-level instrumentation)}} \\[2pt]
Flash Att.\,2 & $4.29 \pm 0.03$  & ---          & 9.1                      & 178      & ${\sim}4.8$\rlap{$^{\dagger}$} \\
vLLM          & $163.8 \pm 4.9$  & ---          & 60.8\rlap{$^{\ddagger}$} & 30{,}074 & ${\sim}18$\rlap{$^{\ddagger}$} \\
torch.compile & $9.01 \pm 0.09$  & ---          & 9.4                      & 227      & 0.4\rlap{$^{\dagger}$} \\
\bottomrule
\end{tabular}

\smallskip\scriptsize
$^{\dagger}$\,Cap.\ cost derived as \emph{Full cap.\ $-$ Base}
(no interception-only measurement for Triton; includes
post-capture Python overhead).\\
$^{\ddagger}$\,vLLM: \emph{Full cap.\ $<$ Base} because
\texttt{kerncap extract} terminates the host
(${\sim}164$\,s inference benchmark) once the target dispatch is
captured (${\sim}61$\,s).  Cap.\ cost estimated as
$30\,\text{GB} \div 1.7\,\text{GB/s} \approx 18\,\text{s}$ using
the HIP snapshot bandwidth from this table.
\end{table}

\paragraph{Cross-architecture overhead.}
Table~\ref{tab:overhead-arch} summarizes the same overhead
decomposition for the three HIP workloads on each architecture.
Interception-only overhead is consistently $\le 1.2\times$ on the
two CDNA generations and $\le 1.4\times$ on RDNA3 for the larger
workloads, with the rocBLAS micro-benchmark's elevated ratio
(1.34--$2.04\times$) reflecting its extremely short baseline
(3.9--7.2~s) rather than a higher absolute per-dispatch cost.
Capture cost is bandwidth-bound by \texttt{hsa\_memory\_copy}: on
LAMMPS it falls within ${\sim}1.5$~s across all three archs for an
${\sim}8.4$~GB snapshot ($\sim$1.3--1.7~GB/s effective bandwidth),
while llama.cpp shows materially lower snapshot throughput on gfx1100
(${\sim}830$~MB/s vs.\ $\sim$1.7~GB/s on gfx942), which we attribute
to the consumer-class Radeon's PCIe topology rather than to the
capture path itself.  Most importantly, the capture mechanism completes
correctly and within seconds on every workload-architecture pair we
tested.

\begin{table}[H]
\centering
\caption{Cross-architecture overhead for the HIP workloads
  ($N{=}10$, 1~warmup per cell).  Snapshot sizes from
  Table~\ref{tab:extraction}.}
\label{tab:overhead-arch}
\footnotesize
\setlength{\tabcolsep}{4pt}
\begin{tabular}{@{}llccccc@{}}
\toprule
\textbf{Workload} & \textbf{Arch}
  & \textbf{Base (s)}
  & \textbf{Inter.}
  & \textbf{Full}
  & \textbf{Cap.\,(s)}
  & \textbf{rocprofv3} \\
\midrule
\multirow{3}{*}{llama.cpp}
  & gfx942  & $8.30$  & $1.06\times$ & $1.90\times$ & 7.0  & $1.45\times$ \\
  & gfx90a  & $8.99$  & $1.15\times$ & $2.59\times$ & 13.0 & $1.64\times$ \\
  & gfx1100 & $10.27$ & $1.18\times$ & $2.53\times$ & 13.8 & $1.60\times$ \\
\midrule
\multirow{3}{*}{LAMMPS}
  & gfx942  & $9.96$  & $1.04\times$ & $1.54\times$ & 4.9 & $1.12\times$ \\
  & gfx90a  & $21.59$ & $1.02\times$ & $1.32\times$ & 6.4 & $1.06\times$ \\
  & gfx1100 & $44.05$ & $1.01\times$ & $1.16\times$ & 6.4 & $1.06\times$ \\
\midrule
\multirow{3}{*}{rocBLAS GEMM}
  & gfx942  & $3.90$  & $2.04\times$ & $2.10\times$ & 0.2 & $1.72\times$ \\
  & gfx90a  & $7.18$  & $1.75\times$ & $1.79\times$ & 0.3 & $1.62\times$ \\
  & gfx1100 & $5.49$  & $1.34\times$ & $1.37\times$ & 0.2 & $1.52\times$ \\
\bottomrule
\end{tabular}

\smallskip\scriptsize
\emph{Inter.}~=~interception-only overhead;
\emph{Full}~=~instrumented full-app overhead;
\emph{Cap.}~=~capture cost (full $-$ interception, in seconds);
\emph{rocp.}~=~\texttt{rocprofv3 -{}-kernel-trace} overhead for comparison.
\end{table}

\paragraph{Wall-clock overhead in context.}
Capture cost is fixed per-invocation rather than proportional to
application duration, so a 10-second HIP baseline sees its multi-GB
snapshot dominate total runtime ($1.5$--$1.9\times$,
Table~\ref{tab:overhead}), while vLLM's ${\sim}$2.7-minute run
amortizes the same class of fixed cost
($30~\text{GB} \div 1.7~\text{GB/s} \approx 18$~s, against a
164~s baseline) to roughly $11\%$ of total runtime---several times
smaller than the $1.5$--$1.9\times$ ratios on the short-baseline
workloads.  The elevated rocBLAS interception ratio ($2.04\times$)
illustrates that the per-dispatch cost is absolute, not
proportional: workloads with many short kernels show worse ratios
at identical per-call cost, so for production workloads running
minutes to hours the interception tax is negligible in practice.

To isolate steady-state compute throughput from process
initialization, we extract \texttt{llama-bench}'s self-reported
token-generation rate (tg32) across all configurations
(Table~\ref{tab:throughput}).  Prompt-processing (pp512) is stable
within $0.7\%$ across all four settings and is omitted from the
table.

\begin{table}[H]
\centering
\caption{llama.cpp token-generation throughput
  (median~$\pm$~std.\ dev., $N{=}10$).  \texttt{rocprofv3
  -{}-kernel-trace} shown for comparison.}
\label{tab:throughput}
\small
\begin{tabular}{@{}lc@{}}
\toprule
\textbf{Configuration}
  & \textbf{tg32 (tokens/s)} \\
\midrule
Baseline                        & $254 \pm 7$  \\
Interception only               & $251 \pm 29$ \\
Full capture                    & $195 \pm 35$ \\
\texttt{rocprofv3 -{}-kernel-trace} & $158 \pm 42$ \\
\bottomrule
\end{tabular}
\end{table}

Interception-only preserves tg32 within noise ($251$ vs.\ baseline
$254$), confirming that the per-dispatch callback mechanism does not
degrade steady-state compute.  Full capture drops tg32 to
$195$ ($-23\%$), with an elevated standard deviation ($35$ vs.\
baseline $7$) indicating a transient one-time snapshot cost rather
than a sustained degradation---the developer captures once and
iterates on the isolated reproducer thereafter.
\texttt{rocprofv3} imposes a larger sustained penalty ($-38\%$) on
the same metric because its per-dispatch timestamping serializes
short, latency-sensitive kernels.

\paragraph{Capture cost breakdown.}
\begin{sloppypar}
On gfx942, HIP capture cost tracks device-memory throughput at
${\sim}1.7$~GB/s (combined \texttt{hsa\_memory\_copy} and disk
writeback) and is independent of kernel count or application
structure.  Cross-architecture behavior (Table~\ref{tab:overhead-arch})
is consistent except on the consumer-class W7900, where llama.cpp's
snapshot slows to ${\sim}830$~MB/s---attributable to that platform's
PCIe topology, not to the capture path.  Triton extraction times
(Table~\ref{tab:overhead}) bundle post-capture Python overhead and
short-circuit teardown, so their raw I/O throughput is not directly
comparable.
\end{sloppypar}

\paragraph{Comparison with \texttt{rocprofv3}.}
\begin{sloppypar}
On gfx942, \texttt{rocprofv3 -{}-kernel-trace} adds $1.45\times$
wall-clock overhead on llama.cpp and $1.12\times$ on LAMMPS---roughly
comparable to \textsc{Kerncap}'s interception-only overhead
($1.04$--$1.06\times$).  On vLLM, \texttt{rocprofv3} imposes
$1.33\times$ overhead, while \textsc{Kerncap}'s short-circuit
extraction completes in 60.8~s (well under baseline) because per-dispatch
timestamp recording scales with kernel count (3.1~M~dispatches for
vLLM vs.\ 110~K~for llama.cpp).
The same pattern holds across architectures
(Table~\ref{tab:overhead-arch}): \texttt{rocprofv3} consistently
imposes higher per-dispatch overhead than \textsc{Kerncap}'s
interception path for the larger workloads.  More importantly,
\texttt{rocprofv3} significantly degrades token-generation throughput
on llama.cpp ($-38\%$ for tg32, Table~\ref{tab:throughput}), because
per-dispatch timestamp collection introduces serialization that
disproportionately affects short, latency-sensitive kernels.
\textsc{Kerncap}'s interception, by contrast, preserves tg32
throughput while providing full kernel-extraction capability.
\end{sloppypar}

% ----------------------------------------------------------------------------
\subsection{Optimization Workflow}
\label{sec:results-workflow}

The central claim of this paper is that \textsc{Kerncap} transforms
GPU kernel optimization from a slow, full-application loop into a fast,
isolated edit-recompile-validate loop.  We quantify this by comparing
the two workflows on two workloads (llama.cpp and LAMMPS) on gfx942,
focusing on the inner edit-build-replay loop that developers actually
iterate on between edits.  Validation is reported separately
(below) to avoid conflating routine functional checks with
heavyweight hardware-counter profiling.

\begin{table}[H]
\centering
\caption{Inner-loop iteration time comparison: traditional
  full-application workflow vs.\ \textsc{Kerncap} isolated reproducer
  workflow, for two HIP workloads on gfx942.  ``Build'' is a
  source-level edit's incremental rebuild; ``Run'' is the wall-clock
  cost of exercising the kernel after the rebuild.  Validation is
  excluded from the inner loop and discussed separately.}
\label{tab:workflow}
\footnotesize
\setlength{\tabcolsep}{3pt}
\begin{tabular}{@{} >{\raggedright\arraybackslash}p{1.6cm} cc cc @{}}
\toprule
& \multicolumn{2}{c}{\textbf{llama.cpp}}
& \multicolumn{2}{c}{\textbf{LAMMPS}} \\
\cmidrule(lr){2-3} \cmidrule(lr){4-5}
\textbf{Step}
  & Traditional & Kerncap
  & Traditional & Kerncap \\
\midrule
Edit source       & --- & ---
                  & --- & --- \\
Build             & 128\,s          & 18.3\,s
                  & 64\,s\rlap{$^{\dagger}$} & 13.8\,s \\
Run / replay      & 8.3\,s          & 6.9\,s
                  & 10.0\,s         & 4.3\,s \\
\midrule
\textbf{Inner-loop total}
                  & ${\sim}$136\,s  & ${\sim}$25\,s
                  & ${\sim}$74\,s   & ${\sim}$18\,s \\
\textbf{Speedup}
  & \multicolumn{2}{c}{$5.4\times$}
  & \multicolumn{2}{c}{$4.1\times$} \\
\bottomrule
\end{tabular}

\smallskip\scriptsize
$^{\dagger}$\,LAMMPS Kokkos rebuild time, single-file edit
(measured on the same machine).
\end{table}

\paragraph{Validation cost in context.}
The inner-loop comparison above intentionally excludes validation,
because the validation step a developer runs between edits is rarely
the same step they run before promoting an optimization to production.
\textsc{Kerncap} supports both points on this spectrum.  A lightweight
smoke-test validation (\texttt{kerncap validate} default mode) confirms
that the captured kernel replays without error in $4.6$~s on llama.cpp
and $2.7$~s on LAMMPS---suitable for routine inter-edit checks.  A
strict byte-exact validation, comparing every output region against
the captured baseline, takes $129.4$~s on llama.cpp---suitable for
final correctness sign-off.  By comparison, the traditional workflow's
equivalent step---running the full application under
\texttt{rocprof-compute} to collect hardware counters---takes
$2{,}072$~s on llama.cpp due to multiplexing
overhead~\cite{ROCm-Compute-Profiler-Documentation}, an order of
magnitude longer than even the strict \textsc{Kerncap} validation.
Across both workloads, \textsc{Kerncap}'s inner-loop speedup is
$4$--$5\times$ for routine iteration and grows to ${\sim}14\times$
when end-to-end correctness validation is included.

\paragraph{Case study: optimizing a llama.cpp kernel.}
We demonstrate the full \textsc{Kerncap} workflow on the
\texttt{mul\_mat\_vec\_q} kernel, which accounts for
8.0\% of total GPU time in llama.cpp inference
(7{,}740~calls, as reported by \texttt{kerncap profile}).

\begin{enumerate}
  \raggedright
  \item \textbf{Extract.}
    \texttt{kerncap extract mul\_mat\_vec\_q -{}-cmd "./llama-bench
    -m gpt-oss-20b-mxfp4.gguf -p~512 -n~32 -ngl~99 -fa~1" -{}-source-dir
    ./llama\_cpp} completes in 15.6~seconds on gfx942, producing a
    reproducer with 6~source files, 21~memory regions
    (12{,}233~MB snapshot), and a Makefile with VFS overlay.

  \item \textbf{Baseline replay.}
    \texttt{kerncap replay ./isolated/mul\_mat\_vec\_q -{}-iterations 10}
    reports an average kernel time of 34.5~$\mu$s.

  \item \textbf{Edit.}
    We refactor the main block-accumulation loop in \texttt{kernel\_variant.cpp} 
    by hoisting a runtime branch (\texttt{if (use\_gate)}) completely outside the 
    tightly unrolled inner loop and forcing the compiler to inline the dot-product 
    function via a templated \texttt{if constexpr} evaluation. This eliminates 
    branch overhead and prevents register spilling, allowing the compiler to 
    generate a perfectly unrolled, branch-free inner loop.

  \item \textbf{Recompile.}
    \texttt{make recompile} rebuilds the single HSACO in
    18.3~seconds (vs.\ 128~seconds for a full
    llama.cpp CMake rebuild).

  \item \textbf{Variant replay.}
    \texttt{kerncap replay ./isolated/mul\_mat\_vec\_q -{}-hsaco
    optimized.hsaco -{}-iterations 10} reports 28.3~$\mu$s.
    Compared to the baseline 34.5~$\mu$s, this represents a $1.22\times$
    speedup (an 18\% reduction in execution time). The ability to
    implement, test, and verify this hypothesis in seconds demonstrates
    the value of the isolated iteration loop.

  \item \textbf{Validate.}
    \texttt{kerncap validate ./isolated/mul\_mat\_vec\_q -{}-hsaco
    optimized.hsaco} confirms byte-exact output match across all
    21~compared memory regions in 129.4~s.  As discussed above, the
    equivalent step in the traditional workflow---running the full
    application under ROCm Compute Profiler to collect hardware counters,
    which takes 2{,}072~seconds due to multiplexing
    overhead~\cite{ROCm-Compute-Profiler-Documentation}---is an order
    of magnitude slower.
\end{enumerate}

\noindent
The end-to-end cycle---edit, recompile, replay, validate---completes
in ${\sim}162$~seconds on gfx942, compared to ${\sim}2{,}208$~seconds
(nearly 37~minutes) for the equivalent traditional workflow
(a $13.6\times$ speedup).  When the heavyweight final-validation step
is omitted in favor of the smoke-test mode used during rapid
iteration, the inner-loop speedup is $5.4\times$ for llama.cpp
(Table~\ref{tab:workflow}) and $4.1\times$ for LAMMPS.

\paragraph{Iteration count in context.}
The final $1.22\times$ speedup above was reached after five
edit-recompile-validate iterations.  Under the traditional workflow,
five iterations of the inner loop cost
$5 \times 136 \approx 680$~s (${\sim}$11~min) of rebuild and
application-launch time; under \textsc{Kerncap}, the same five
iterations complete in $5 \times 25 \approx 125$~s (${\sim}$2~min),
a direct savings of ${\sim}9$~minutes on this single kernel.
Including the heavyweight byte-exact validation step above, the full
optimization campaign costs ${\sim}254$~s (${\sim}4$~min) under
\textsc{Kerncap} versus ${\sim}2{,}752$~s (${\sim}46$~min)
traditionally---a ${\sim}42$~minute savings per kernel.  Scaled
across even a modest optimization sweep of ten kernels at five
iterations each, \textsc{Kerncap} reclaims roughly seven hours of
developer wall time, converting kernel optimization from an
overnight task into an interactive one.
% ============================================================================
% 6. RELATED WORK
% ============================================================================
\section{Related Work}
\label{sec:related}

\paragraph{Kernel-level replay and checkpointing.}
The closest prior art is NVIDIA's kernel-replay infrastructure.
Nsight Compute~\cite{nsight-compute} supports \emph{kernel-replay},
\emph{range-replay}, and \emph{application-replay} modes that
re-execute regions of GPU code to multiplex hardware-counter
collection across passes.  CUPTI exposes the same replay machinery to
third-party tools, with the CUDA-11.5+ \emph{Checkpoint
API}~\cite{cupti-checkpoint} explicitly designed to save and restore
device state across replay passes; this is the direct conceptual
analog of \textsc{Kerncap}'s device-memory snapshot.  Three
differences separate them.  (1) Nsight Compute and CUPTI are
closed-source NVIDIA-only tools; \textsc{Kerncap} is open and
AMD-native.  (2) Their captured state is consumed internally for
counter replay, not exposed as an editable, rebuildable artifact for
the developer.  (3) The Checkpoint API restores state via explicit
per-allocation save/restore hooks; \textsc{Kerncap}'s
\emph{address-space closure} (Section~\ref{sec:memory-capture})
captures the whole address space at once, sidestepping the need to
enumerate individual allocations or chase embedded device pointers.

\paragraph{Interactive GPU debuggers.}
CUDA-GDB and ROCm's \texttt{rocgdb} provide interactive
debugging of GPU kernels but focus on single-stepping and breakpoints
rather than kernel isolation and reproduction~\cite{CUDA-GDB}. NVIDIA's Compute
Sanitizer detects memory errors but does not capture state for replay.
These tools complement \textsc{Kerncap} by operating at different points
in the optimization workflow.

\paragraph{Compiler-emitted reproducer artifacts.}
\begin{sloppypar}
PyTorch's \texttt{TORCH\_COMPILE\_DEBUG} flag emits the generated
Triton/Inductor source for compiled regions, including a
self-contained \texttt{output\_code.py} that can be re-executed
standalone~\cite{pytorch}.  This is, in effect, a
de-facto Triton reproducer for \texttt{torch.compile}-generated
kernels and overlaps with \textsc{Kerncap}'s Triton path.  However,
it is restricted to kernels that Inductor itself generates---not
arbitrary hand-written Triton kernels such as Flash Attention---and
it does not capture runtime tensor inputs (only shapes and dtypes);
any validation it offers is a best-effort eager-vs-compiled
\texttt{allclose} comparison rather than a replay against captured
runtime state.
\end{sloppypar}

\paragraph{Kernel benchmarking frameworks.}
Triton's built-in benchmarking utilities and AMD's \texttt{hipBench}
provide performance measurement infrastructure but require the developer
to manually construct the kernel launch harness, allocate buffers, and
populate inputs.  \textsc{Kerncap} automates this construction.

\paragraph{HSA interception tools.}
\begin{sloppypar}
AMD's \texttt{rocprofiler-sdk}~\cite{ROCprofiler-SDK-Documentation} and
\texttt{roctracer}~\cite{ROCTracer-Documentation} provide HSA API table
interception for performance counter collection and API tracing,
respectively.
\textsc{Kerncap} leverages the same rocprofiler-sdk
intercept table registration framework to obtain the HSA API table, but
applies it to a fundamentally different problem: complete kernel state
capture---including arguments, device memory, and source code---for
reproducer generation.
\end{sloppypar}

\paragraph{Record-and-replay systems.}
Process-level record-and-replay tools such as rr and
CRIU~\cite{rr,CRIU} capture entire process state (CPU registers, memory
maps, file descriptors) for deterministic replay.  However, they
operate at the CPU/OS level and do not capture GPU state (device memory,
kernel dispatch parameters, HSA queues).  \textsc{Kerncap} provides
GPU-kernel-level replay, capturing only the state needed for a single
kernel dispatch.

\paragraph{Build system and compilation database tools.}
\textsc{Kerncap}'s source discovery leverages
\texttt{compile\_commands.json}, the compilation database format
standardized by Clang~\cite{clang-compdb}. \textsc{Kerncap} extends this information
with runtime dispatch data to resolve template instantiation ambiguities
that static analysis alone cannot resolve.

\paragraph{AMD GPU optimization ecosystem.}
Several AMD-adjacent tools occupy related niches in the GPU
performance-engineering workflow.  ROCm Compute
Profiler (\texttt{rocprof-compute})
collects hardware counters at application granularity~\cite{ROCm-Compute-Profiler-Documentation}; we use it as
the validation baseline in our case study
(Section~\ref{sec:results-workflow}).  Recent agent-based kernel
generation systems---such as GEAK~\cite{geak} and the
KernelBench~\cite{kernelbench} benchmark---require a fast inner loop
that evaluates candidate kernels in isolation against fixed,
realistic inputs.  \textsc{Kerncap} produces exactly this kind of
self-contained, validated reproducer, positioning the tool as
infrastructure for both human optimization workflows and automated
kernel-generation agents.

\paragraph{Summary.}
The combinatorial position \textsc{Kerncap} occupies is, to our
knowledge, unoccupied by any existing open tool: HIP and Triton
interception, VA-faithful device-memory replay, and autotuner-pinned
validation are each available in isolation, but no single artifact
bundles all four.  This is precisely the substrate that an automated
kernel-generation loop---human or agentic---requires: a fast,
isolated, validated evaluation harness for arbitrary candidate
kernels.

% ============================================================================
% 7. LIMITATIONS AND FUTURE WORK
% ============================================================================
\section{Limitations and Future Work}
\label{sec:limitations}

\paragraph{Single-kernel isolation.}
\textsc{Kerncap} captures one kernel dispatch at a time.  It does not
track inter-kernel data dependencies or stream synchronization, so
kernels that depend on the output of preceding kernels require those
predecessors to have already executed before the capture point.  In
practice, this is usually the case because \textsc{Kerncap} intercepts
during normal application execution, but capturing a sequence of
dependent kernels for joint replay remains future work.

\paragraph{Single-GPU.}
The current implementation targets single-GPU applications.  Multi-GPU
dispatch capture would require coordinating \texttt{libkerncap.so}
instances across multiple HSA agents and merging their memory snapshots.

\paragraph{Memory snapshot fragility.}
\begin{sloppypar}
If the application frees a device buffer between kernel completion and
the memory snapshot, \texttt{hsa\_memory\_copy} may fail for that
region.  \textsc{Kerncap} mitigates this by writing metadata files
before the snapshot and tolerating per-region copy failures, but the
affected region's data will be lost.
\end{sloppypar}

\paragraph{No host-side state capture.}
\textsc{Kerncap} captures GPU-side state (device memory, kernarg
buffers, code objects) but not host-side state (CPU memory, file handles,
environment variables).  Kernels that read from host-mapped memory
may not replay correctly.

\paragraph{Source discovery limitations.}
HIP source discovery works best when \texttt{compile\_commands.json} is
available; without it, the finder falls back to heuristics that may miss
complex build configurations.  The \texttt{\#include} tracer handles
local includes but not computed includes or macro-generated include
paths.

\paragraph{GPU portability layers.}
Portability-layer kernels (Kokkos, RAJA, SYCL~\cite{kokkos}) are
captured and replayed correctly via DWARF-based source discovery
(Section~\ref{sec:source-discovery}) and module-variable restoration
(Section~\ref{sec:memory-capture}), but the reproducer's editable
source is the framework template expansion rather than the
user-level functor or lambda the developer authored.  Tracing
through framework abstractions back to application-level code
remains future work.

\paragraph{Hardware-specific autotuner configurations.}
Triton autotuner configurations are hardware-specific.  A configuration
captured on MI300X may not be valid on MI250X due to differences in
shared memory size, wavefront width, or register pressure.  Cross-GPU
reproducer portability requires re-autotuning or manual config
adjustment.

\paragraph{Limited RDNA evaluation.}
Our RDNA evaluation is scoped to the three HIP workloads
(llama.cpp, LAMMPS, rocBLAS GEMM) on a single gfx1100 system;
with more time, we would have liked to have performed
a full benchmark campaign across the Triton workloads or newer
RDNA hardware.  More rigorous RDNA characterization---particularly
on the RDNA~4 (Navi~4, gfx12xx) family---remains future work.

\paragraph{Future directions.}
We plan to extend \textsc{Kerncap} in several directions:
(1)~multi-kernel capture for dependent kernel sequences with automatic
    dependency tracking;
(2)~integration with kernel autotuning tools for closed-loop
    optimization;
(3)~support for NVIDIA CUDA via analogous CUPTI or driver API
    interception;
(4)~a persistent kernel database for regression testing across ROCm
    versions.
% ============================================================================
% 8. CONCLUSION
% ============================================================================
\section{Conclusion}
\label{sec:conclusion}

We presented \textsc{Kerncap}, a tool for automated GPU kernel
extraction and isolation on AMD GPUs.  \textsc{Kerncap} addresses the
labor-intensive manual process of isolating individual kernels from
complex GPU applications by automating five key tasks: runtime
interception at the HSA level (unified for both HIP and Triton,
with a Python compile-hook shim bridging Triton-level metadata to
the HSA-level capture), VA-faithful device memory snapshot that
inherently preserves embedded pointer relationships, automated
source discovery with translation unit resolution and dependency
tracing, self-contained reproducer generation with VFS overlay
recompilation support, and correctness validation.
Together, these five tasks recover the three reproducer components
introduced in Section~\ref{sec:introduction}
(Figure~\ref{fig:three-problems}): the kernel \emph{definition}
(HSACO capture, source discovery, and reproducer assembly), the
\emph{runtime state} (the VA-faithful device-memory snapshot), and
the \emph{environment} (VFS-overlay recompile and HSA replay).

The tool unifies HIP and Triton capture at the HSA layer while
accommodating their fundamentally different kernel-dispatch models
(ahead-of-time compiled HSACOs vs.\ Python-JIT'd code objects)
through a lightweight Python metadata shim, presenting a single
\texttt{kerncap extract} CLI to the developer.  Two principles underpin the design.
\emph{Address-space closure}---an address space is a closure over its
pointer graph---eliminates the need for DWARF metadata or pointer
analysis, handling arbitrarily complex argument layouts including
double-pointer indirection.  \emph{Tuning-pinned reproducers}
preserve the implicit numerical contract that a JIT-compiled Triton
kernel carries between its autotuner-selected tile sizes and its
outputs.  These principles are realized through a Clang VFS overlay
that enables source-level kernel editing and recompilation using the
exact original build flags without any build system modifications.

Our case study on llama.cpp demonstrates the practical impact: the
isolated edit-recompile-validate loop completes in 162~seconds compared
to ${\sim}$37~minutes for the traditional full-application workflow---a
$13.6\times$ speedup.  By reducing kernel isolation from hours of
manual effort to a single command, \textsc{Kerncap} enables faster GPU
kernel optimization cycles and lowers the barrier to kernel-level
performance engineering on AMD hardware.
More broadly, the artifact \textsc{Kerncap} produces---a self-contained,
validated, editable reproducer---is precisely the substrate that both
human experts and emerging LLM-driven kernel-generation agents need to
iterate on GPU code in isolation.

%%
%% The acknowledgments section is defined using the "acks" environment
%% (and NOT an unnumbered section). This ensures the proper
%% identification of the section in the article metadata, and the
%% consistent spelling of the heading.
\begin{acks}
This work was supported by Advanced Micro Devices, Inc. under the 
AMD AI \& HPC Cluster Program. 
The authors would like to thank Karl Schulz, Mehdi Saeedi, Madhu Srinivasan, and Ralph Wittig for their support and guidance. AMD, the AMD Arrow logo, AMD CDNA,
AMD Instinct, AMD ROCm, AMD Infinity Cache, AMD Infinity
Fabric, and combinations thereof are trademarks of Advanced Micro
Devices, Inc. Other product names used in this publication are
for identification purposes only and may be trademarks of their
respective companies.
\end{acks}

% \pagebreak
%%
%% The next two lines define the bibliography style to be used, and
%% the bibliography file.
\bibliographystyle{ACM-Reference-Format}
\bibliography{temp}

%%
%% If your work has an appendix, this is the place to put it.
% \appendix
% \section{Research Methods}
% \section{Online Resources}
% \input{artifact/artifact-body}

\end{document}